\documentclass[aps,preprint]{revtex4}
\usepackage{amssymb}

\usepackage{latexsym}
\usepackage{amsfonts}
\usepackage{graphicx}
\usepackage{epsfig}
\usepackage{amsmath}

\begin{document}

\title{The Growth of Business Firms:\\
Theoretical Framework and Empirical Evidence}

\author{Dongfeng~Fu$^{\ast }$, Fabio~Pammolli$^{\ast\ddag\S }$,
S.~V.~Buldyrev$^{\P }$, Massimo~Riccaboni$^{\ddag\S }$, Kaushik~Matia$^{\ast
}$, Kazuko~Yamasaki$^{\parallel}$, H.~E.~Stanley$^{\ast}$ }


\affiliation{$^{\ast}$Center for Polymer Studies and Department of Physics,
 Boston University, Boston, MA 02215 USA \\$^{\ddag }$Faculty of Economics,
 University of Florence, Milan, Italy \\$^{\S}$IMT Institute for Advanced
 Studies, Via S.~Micheletto 3, Lucca, 55100 Italy\\ $^{\P}$Department of
 Physics,~Yeshiva University, 500 West 185th Street,~New York, NY 10033 USA\\
 $^{\parallel}$Tokyo University of Information Sciences, Chiba City 265-8501
 Japan}

\begin{abstract}
We introduce a model of proportional growth to explain the distribution
$P_g(g)$ of business firm growth rates. The model predicts that $P_g(g)$ is
exponential in the central part and depicts an asymptotic power-law behavior
in the tails with an exponent $\zeta=3$. Because of data limitations,
previous studies in this field have been focusing exclusively on the Laplace
shape of the body of the distribution. In this article, we test the model at
different levels of aggregation in the economy, from products to firms to
countries, and we find that the model's predictions agree with empirical
growth distributions and size-variance relationships.
\end{abstract}


\maketitle

\section{Introduction}
\label{sec:Introduction}
Gibrat~\cite{Gibrat30, Gibrat31}, building upon the work of the astronomers
Kapteyn~\cite{Kapteyn16}, assumed the expected value of the growth rate of a
business firm's size to be proportional to the current size of the firm,
which is called ``Law of Proportionate Effect''~\cite{Zipf49,Gabaix99}. Several
models of proportional growth have been subsequently introduced in economics
in order to explain the growth of business firms~\cite{Steindl65, Sutton97,
Kalecki45}. Simon and co-authors~\cite{Simon55, Simon58, Simon75, Simon77}
extended Gibrat's model by introducing an entry process according to which
the number of firms rise over time. In Simon's framework, the market consists of a
sequence of many independent ``opportunities'' which arise over time, each of
size unity. Models in this tradition have been challenged by many
researchers~\cite{Stanley96, Lee98, Stanley99, Bottazzi01, Matia04} who found
that the firm growth distribution is not Gaussian but displays a tent shape.

Here we introduce a general framework that provides an unifying explanation
for the growth of business firms based on the number and size distribution of
their elementary constituent
components~\cite{Amaral97,Sergey_II,Sutton02,DeFabritiis03,Amaral98,Takayasu98,Canning98,
Buldyrev03}. Specifically we present a model of proportional growth in both
the number of units and their size and we draw some general implications on
the mechanisms which sustain business firm growth~\cite{Simon75, Sutton97,
Kalecki,Mansfield,Hall,DeFabritiis03}.  According to the model, the
probability density function (PDF) of growth rates is Laplace in the
center~\cite{Stanley96} with power law tails~\cite{Reed01,Reed02} decaying as
$P_g(g)\thicksim g^{-\zeta }$ where $\zeta=3$.

Because of data limitations, previous studies in this field focus on the
Laplace shape of the body of the distribution~\cite{Kotz01}. Using a database
on the size and growth of firms and products, we characterize the shape of
the whole growth rate distribution.

We test our model by analyzing different levels of aggregation of economic
systems, from the ``micro'' level of products to the ``macro'' level of
industrial sectors and national economies. We find that the model accurately
predicts the shape of the PDF of growth rate at all levels of aggregation
studied.

\section{The Theoretical Framework}
\begin{figure}[tbp]
\centering
\includegraphics[width=8cm,angle=-90]{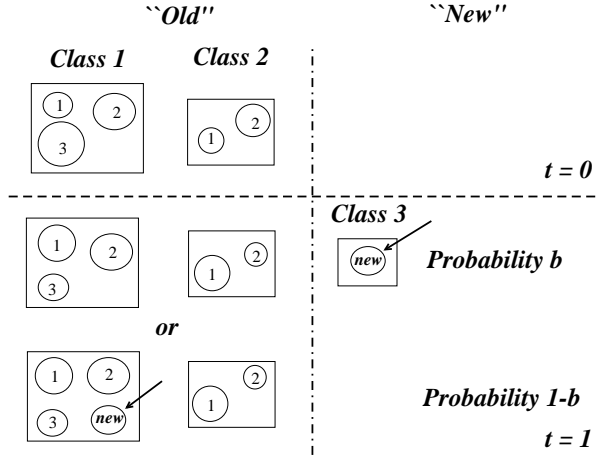}
\caption{Schematic representation of the model of proportional growth. At
time $t=0$, there are $N(0)=2$ classes ($\square$) and $n(0)=5$ units
($\bigcirc$) (Assumption A1). The area of each circle is proportional to the
size $\xi$ of the unit, and the size of each class is the sum of the areas of
its constituent units (see Assumption B1). At the next time step, $t=1$, a new unit is
created (Assumption A2). With probability $b$ the new unit is assigned to a
new class (class 3 in this example) (Assumption A3). With probability $1-b$
the new unit is assigned to an existing class with probability proportional
to the number of units in the class (Assumption A4). In this example, a new
unit is assigned to class $1$ with probability $3/5$ or to class $2$ with
probability $2/5$. Finally, at each time step, each circle $i$ grows or
shrinks by a random factor $\eta_i$ (Assumption B2).}
\label{schematic}
\end{figure}

We model business firms as classes consisting of a random number of units.
According to this view, a firm is represented as the aggregation of its
constituent units such as divisions~\cite{Amaral98},
businesses~\cite{Sutton02}, or products~\cite{DeFabritiis03}. Accordingly, on
a different level of coarse-graining, a class can represent a national
economy composed by economic units such as firms. In this article we study the
logarithm of the one-year growth rate of classes $g\equiv \log(S(t+1)/S(t))$
where $S(t)$ and $S(t+1)$ are the sizes of classes in the year $t$ and $t+1$
measured in monetary values (GDP for countries, sales for firms and
products). Our model is illustrated in Fig.~\ref{schematic}. Two key sets of
assumptions in the model are (A) the number of units in a class grows in
proportion to the existing number of units and (B) the size of each unit
fluctuates in proportion to its size.

The first set of assumptions is:
\begin{itemize}
\item[{(A1)}] Each class $\alpha$ consists of $K_{\alpha}(t)$ number of
  units. At time $t=0$, time step measured by year generally, there are
  $N(0)$ classes consisting of $n(0)$ total number of units. The initial
  average number of units in a class is thus $n(0)/N(0)$.

\item[{(A2)}] At each time step a new unit is created. Thus the number
  of units at time $t$ is $n(t)=n(0)+t$.

\item[{(A3)}] With birth probability $b$, this new unit is assigned to a new
  class, so that the average number of classes at time $t$ is $N(t)=N(0)+bt$.

\item[{(A4)}] With probability $1-b$, a new unit is assigned to an existing
  class $\alpha$ with probability $P_{\alpha}=(1-b)K_{\alpha}(t)/n(t)$, so
  $K_{\alpha}(t+1) = K_{\alpha}(t)+1$.
\end{itemize}

For simplicity, we do not consider the decrease of the number of units in a
class. In reality, elementary units enter and exit. Because we are considering
the case of a growing economy, it is legitimate to assume the entry rate
being higher than the exit rate. On the average, the net entry rate of units
can be simplified as a positive constant. In the model, the net entry rate of
units is fixed at $1$. Thus, at large $t$, it gives results equivalent to the ones
that would have been obtained considering a value for the exit rate of units.

Our goal is to find $P(K)$, the probability distribution of the number of
units in the classes at large $t$. This model in two limiting cases (i)
$b=0$, $K_{\alpha}=1$ $(\alpha =1,2 \ldots N(0))$ and (ii) $b\neq 0$,
$N(0)=1$, $n(0)=1$ has exact analytical solutions
$P(K)=N(0)/t(t/(t+N(0)))^K (1+O(1/t))$~\cite{Johnson,Kotz2000} and
$\lim\limits_{t\to\infty}P(K)=(1+b)\Gamma(K)
\Gamma(2+b)/\Gamma(K+2+b)$~\cite{Reed04} respectively.


In the general case, the exact analytical solution is not known and we obtain
a numerical solution by computer simulations and compare it with the
approximate mean field solution. (see, e.g., Chapter 6 of~\cite{book} and
Appendix A)

Our results are consistent with the exactly solvable limiting cases as well
as with the empirical data on the number of products in the pharmaceutical
firms and can summarized as follows. In the limit of large $t$, the
distribution of $K$ in the old classes that existed at $t=0$ converges to an
exponential distribution~\cite{Cox}
\begin{equation}
P_{\rm old}(K)=\lambda^K{1\over K(t)-1}\approx {1\over K(t)}\exp(-K/K(t)),
\label{P_K_old}
\end{equation}
where $\lambda=1-1/K(t)$ and $K(t)$ is the average number of units in the old
classes at time $t$, $K(t)=[(n(0)+t)/n(0)]^{1-b}\cdot n(0)^{b}/N(0)$. The
distribution of units in the new classes created at $t>0$ converges to a
power law with an exponential cutoff
\begin{equation}
P_{\rm new}(K)\sim K^{-\varphi}f(K),
\label{P_K_new}
\end{equation}
where $\varphi\approx 2+b$ for small $b$, and $f(K)$ decays for $K\to\infty$
faster than $P_{\rm old}(K)$. The distribution of units in all classes
is given by
\begin{equation}
P(K)= \frac{N(0)}{N(0) + bt}\,P_{old}(K)\,+\,\frac{bt}{N(0)+bt}\,P_{new}(K).
\label{p_K_final}
\end{equation}
The mean field approximation for $P_{new}(K)$ is given by 
\begin{eqnarray}  \label{new_PK}
{P}_{new}(K) & \approx & \frac{n(0)/t +1}{1-b}\,K^{\left(-1/(1-b) -
1\right)}\,\int^K_{K^{\prime}} e^{-y}
\,\,y^{\frac{1}{1-b}} \,\,dy.
\label{P_new_K_full}
\end{eqnarray}
where $K^{\prime}=K[n(0)/(n(0)+t)]^{1-b}$.

The second set of assumptions is:

\begin{itemize}
\item[{(B1)}] At time $t$, each class $\alpha$ has $K_{\alpha}(t)$ units of
size $\xi_i(t)$, ${i=1,2,...K_{\alpha}(t)}$ where $K_{\alpha}$ and $\xi_i >
0$ are independent random variables taken from the distributions
$P(K_{\alpha})$ and $P_{\xi}(\xi_i)$ respectively. $P(K_{\alpha})$ is defined
by Eq.~(\protect\ref{p_K_final}) and $P_{\xi}(\xi_i)$ is a given distribution
with finite mean and standard deviation and $\ln\xi_i$ has finite mean
$\mu_{\xi}=\langle\ln\xi_i\rangle$ and variance
$V_{\xi}=\langle(\ln\xi_i)^2\rangle-\mu_{\xi}^2$. The size of a class is
defined as $S_{\alpha}(t)\equiv \sum_{i=1}^{K_{\alpha}}\xi_i(t)$.
\item[{(B2)}] At time $t+1$, the size of each unit is decreased or
  increased by a random factor $\eta_i(t)>0$ so that
\begin{equation}
\xi_i(t+1)=\xi_i(t)\,\eta_i(t),
\end{equation}
where $\eta_i(t)$, the growth rate of unit $i$, is independent random
variable taken from a distribution $P_\eta(\eta_i)$, which has a finite mean.
We also assume that $\ln \eta_i$ has finite mean
$\mu_{\eta}\equiv\langle\ln\eta_i\rangle$ and variance
$V_{\eta}\equiv\langle(\ln\eta_i)^2\rangle-\mu_{\eta}^2$.
\end{itemize}

The growth rate of each class is defined as
\begin{equation}
g_{\alpha}\equiv \log\left(\frac{S_{\alpha}(t+1)}{S_{\alpha}(t)}\right) =
\log\sum_{i=1}^{K_{\alpha}}\xi_i(t+1)-\log\sum_{i=1}^{K_{\alpha}}\xi_i(t).
\end{equation}

Here we neglect the influx of the new units, so
$K_{\alpha}=K_{\alpha}(t+1)=K_{\alpha}(t)$. The resulting distribution of the
growth rates of all classes is determined by
\begin{equation}
P_g(g) \equiv \sum_{K=1}^{\infty}P(K)P_g(g|K),
\label{P_g_g_sum}
\end{equation}
where $P(K)$ is the distribution of the number of units in the classes,
computed in the previous stage of the model and $P_g(g|K)$ is the conditional
distribution of growth rates of classes with given number of units determined
by the distribution $P_{\xi}(\xi)$ and $P_{\eta}(\eta)$.

The analytical solution of this model can be obtained only for certain
limiting cases but a numerical solution can be easily computed for any set of
assumptions. We investigate the model numerically and analytically (see
Appendix B) and find:

\begin{itemize}

\item[{(1)}] The conditional distribution of the logarithmic growth
  rates $P_g(g|K)$ for the firms consisting of a fixed number of units
  converges to a Gaussian distribution for $K\to\infty$:
\begin{equation}
P_g(g|K)\approx{\sqrt K\over\sqrt{2\pi V_g}}\,\exp\left((g-\bar
g)^2K/2V_g\right),
\label{P_g_gk}
\end{equation}
where $V_g$ is a function of parameters of the distribution $P_{\xi}(\xi)$
and $P_{\eta}(\eta)$, and $\bar g$ is mean logarithmic growth rate of a unit,
$\bar g=\langle\ln\eta_i\rangle$.

Thus the width of this distribution decreases as $1/\sqrt K$. This result is
consistent with the observation that large firms with many production units
fluctuate less than small firms~\cite{Sutton97,Amaral97,Amaral98,Hymer}.

\item[{(2)}] For $g\gg V_{\eta}$, the distribution
  $P_g(g)$ coincides with the distribution of the logarithms of the growth
  rates of the units:
\begin{equation}
P_g(g)\approx P_\eta(\ln\eta).
\end{equation}
In the case of power law distribution $P(K)\sim K^{-\varphi}$ which dramatically
  increases for $K\to 1$, the distribution $P_g(g)$ is dominated by the growth
  rates of classes consisting of a single unit $K=1$, thus the distribution
  $P_g(g)$ practically coincides with $P_{\eta}(\ln\eta_i)$ for all $g$. Indeed,
  our empirical observations confirm this result. 

\item[{(3)}] If the distribution $P(K)\sim K^{-\varphi}$, $\varphi>2$ for
  $K\to\infty$, as happens in the presence of the influx of new units $b\neq
  0$, $P_g(g)=C_1-C_2|g|^{2\varphi-3}$, for $g\to 0$ which in the limiting case
  $b\to 0$, $\varphi\to 2$ gives the cusp $P_g(g)\sim C_1-C_2|g|$ ($C_1$ and
  $C_2$ are positive constants), similar to the behavior of the Laplace
  distribution $P_{\rm L}(g)\sim \exp(-|g|C_2)$ for $g\to 0$.

\item[{(4)}] If the distribution $P(K)$ weakly depends on $K$ for $K\to 1$,
  the distribution of $P_g(g)$ can be approximated by a power law of $g$:
  $P_g(g)\sim |g|^{-3}$ in wide range $\sqrt{V_g/K(t)}\ll g\ll\sqrt{V}$, where
  $K(t)$ is the average number of units in a class. This case is realized for
  $b=0$, $t\to\infty$ when the distribution of $P(K)$ is dominated by the
  exponential distribution and $K(t)\to\infty$ as defined by
  Eq.~(\protect\ref{P_K_old}). In this particular case, $P_g(g)$ for $g\ll
  \sqrt{V_g}$ can be approximated by
\begin{equation}
P_g(g) \approx \frac{\sqrt{K(t)}}{2\sqrt{2V_g}}\left(1 + \frac{K(t)}{2V_g}
g^2 \right)^{-3/2}.
\label{g_ll_V}
\end{equation}
\item[{(5)}] In the case in which the distribution $P(K)$ is not dominated by
one-unit classes but for $K\to\infty$ behaves as a power law, which is the
result of the mean field solution for our model when $t\to\infty$, the
resulting distribution $P_g(g)$ has three regimes, $P_g(g)\sim
C_1-C_2|g|^{2\varphi-3}$ for small $g$, $P_g(g)\sim |g|^{-3}$ for intermediate $g$,
and $P_g(g)\sim P(\ln\eta)$ for $g\to\infty$. The approximate solution of
$P_g(g)$ is obtained by using Eq.~(\ref{P_g_gk}) for $P_g(g|K)$ for finite
$K$, mean field solution Eq.~(\ref{P_new_K_full}) in the limit $t\to\infty$
for $P(K)$ and replacing summation by integration in Eq.~(\ref{P_g_g_sum}):
\begin{equation}
P_{g}(g) = {\frac{1}{1-b}}\,{\frac{1}{\sqrt{2\pi
V}}}\int_0^\infty\,\exp(-y) \,y^{\frac{1}{1-b}}\,dy\,\int_y^\infty\,
\exp(-g^2\,K/2V)\,K^{(-\frac{1 }{2}-\frac{1}{1-b}) }\,dK.  \label{b_ne_0}
\end{equation}
For $b\neq 0 $ the integral above can be expressed in elementary
functions. In the $b\to 0$ case, Eq.~(\protect\ref{b_ne_0}) yields the main
result
\begin{equation}
P_g(g) \approx \frac{2V_g}{\sqrt{g^2+2V_g}\,(|g|+\sqrt{g^2+2V_g})^2},\,\,\,\,\,\,\,\,\,\,\,\,\,\, (b\to 0)
\label{p_new_1}
\end{equation}
which combines the Laplace cusp for $g\to 0$ and the power law decay
$|g|^{-3}$ for $g\to\infty$. Note that due to replacement of summation by
integration in Eq.~(\ref{P_g_g_sum}), the approximation Eq.~(\ref{p_new_1})
holds only for $g<\sqrt{V_{\eta}}$.
\end{itemize}  

\begin{figure}[tbp]
\centering
\includegraphics[width=8cm,angle=-90]{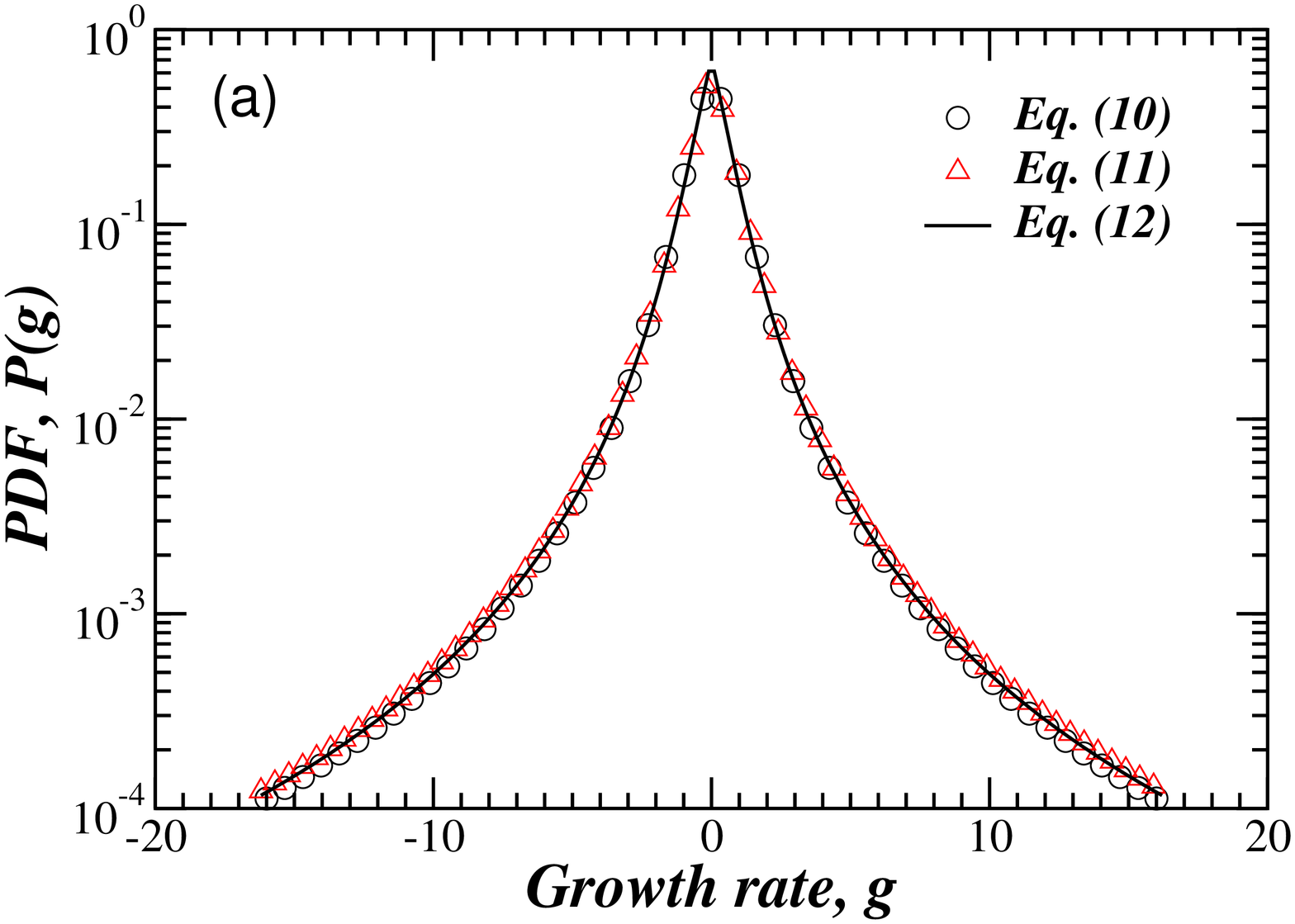}
\includegraphics[width=8cm,angle=-90]{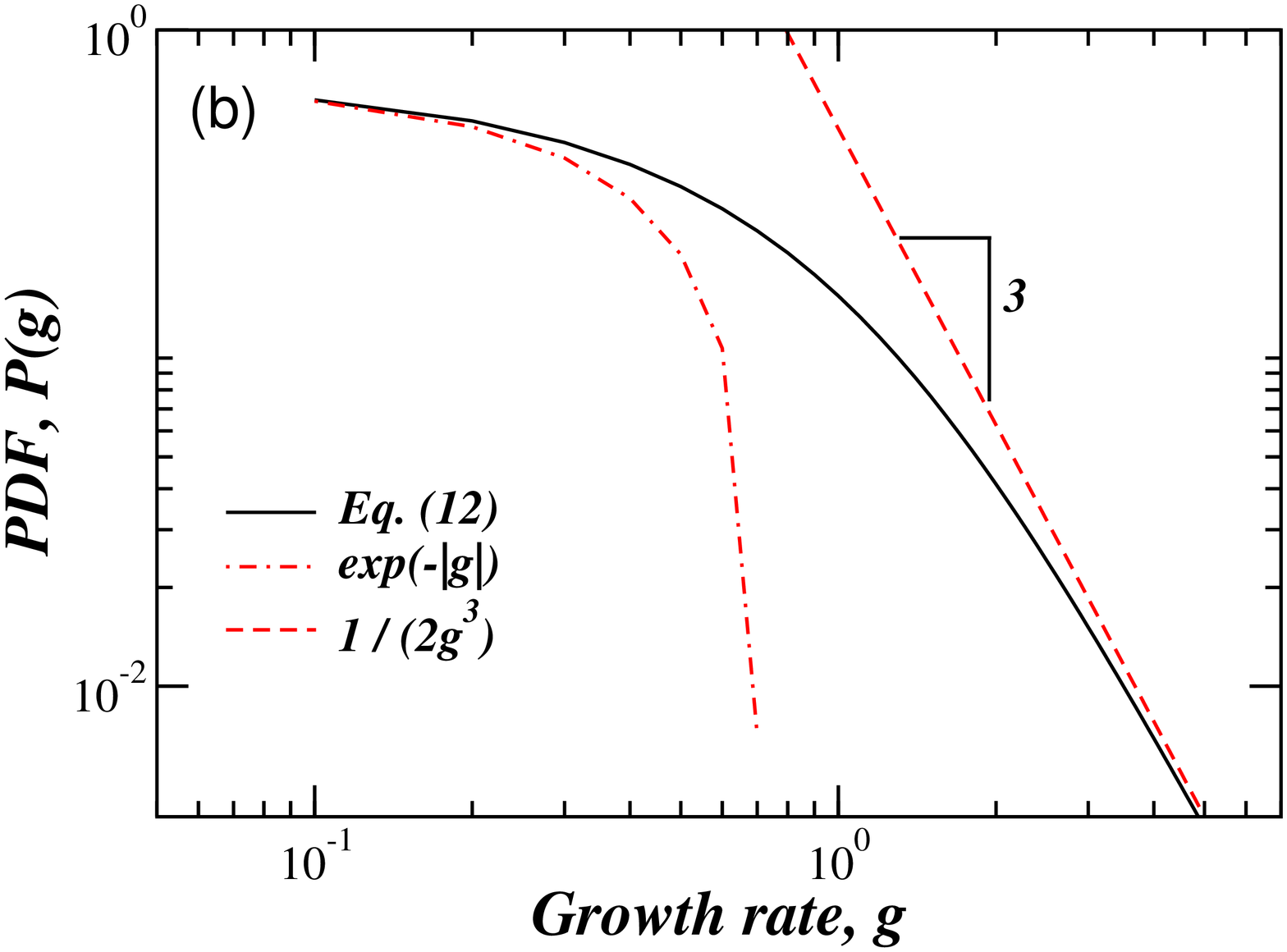}
\caption{ (a) Comparison of three different approximations for the growth
rate PDF, $P_g(g)$, given by Eq.~(\protect\ref{g_ll_V}), mean field
approximation Eq.~(\protect\ref{b_ne_0}) for $b=0.1$ and
Eq.~(\protect\ref{p_new_1}). Each $P_g(g)$ shows similar tent shape behavior
in the central part. We see there is little difference between the three
cases, $b=0$ (no entry), $b=0.1$ (with entry) and the mean field
approximation. This means that entry of new classes ($b>0$) does not
perceptibly change the shape of $P_g(g)$. Note that we use $K(t)/V_g = 2.16$
for Eq.~(\protect\ref{g_ll_V}) and $V_g=1$ for
Eq.~(\protect\ref{p_new_1}). (b) The crossover of $P_g(g)$ given by
Eq.~(\protect\ref{p_new_1}) between the Laplace distribution in the center
and power law in the tails. For small $g$, $P_g(g)$ follows a Laplace
distribution $P_g(g) \sim \exp(-|g|)$, and for large $g$, $P_g(g)$
asymptotically follows an inverse cubic power law $P_g(g) \sim g^{-3}$.}
\label{crossover}
\end{figure}

 In Fig.~\ref{crossover}a we compare the distributions given by
 Eq.~(\protect\ref{g_ll_V}), the mean field approximation
 Eq.~(\protect\ref{b_ne_0}) for $b=0.1$ and Eq.~(\protect\ref{p_new_1}) for
 $b\to 0$. We find that all three distributions have very similar tent shape
 behavior in the central part. In Fig.~\ref{crossover}b we also compare the
 distribution Eq.~(\protect\ref{p_new_1}) with its asymptotic behaviors for
 $g\to 0$ (Laplace cusp) and $g\to \infty$ (power law), and find the
 crossover region between these two regimes.

\section{The Empirical Evidence}
To test our model, we analyze different levels of aggregation of economic
systems, from the micro level of products to the macro level of industrial
sectors and national economies.

First, we analyze a new and unique database, the pharmaceutical industry
database (PHID), which records sales figures of the 189,303 products
commercialized by 7,184 pharmaceutical firms in 21 countries from 1994 to
2004, covering the whole size distribution for products and firms and
monitoring the flows of entry and exit at both levels kindly provided by the
EPRIS program. Then, we study the growth rates of all U.S.  publicly-traded
firms from 1973 to 2004 in all industries, based on Security Exchange
Commission filings (Compustat). Finally, at the macro level, we study the
growth rates of the gross domestic product (GDP) of 195 countries from 1960
to 2004 (World Bank).

\begin{figure}[tbp]
\centering
\includegraphics[width=10cm,angle=-90]{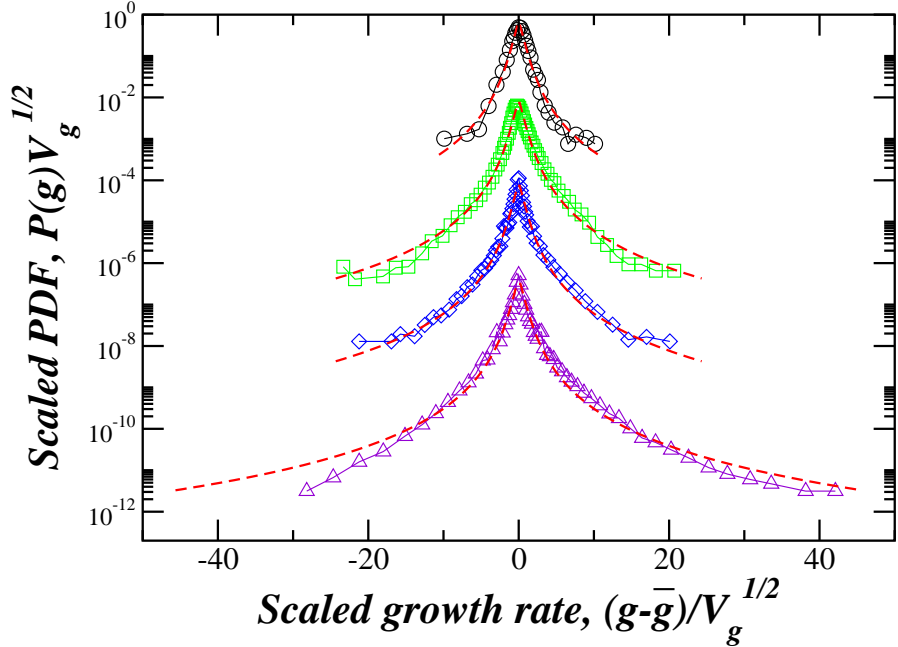}
\caption{Empirical tests of Eq.~(\protect\ref{p_new_1}) for the probability
density function (PDF) $P_g(g)$ of growth rates rescaled by
$\sqrt{V_g}$. Shown are country GDP ($\bigcirc$), pharmaceutical firms
($\square$), manufacturing firms ($\Diamond$), and pharmaceutical products
($\bigtriangleup$). The shapes of $P_g(g)$ for all four levels of aggregation
are well approximated by the PDF predicted by the model (dashed
lines). Dashed lines are obtained based on Eq.~(\protect\ref{p_new_1}) with
$V_g\approx 4\times10^{-4}$ for GDP, $V_g\approx 0.014$ for pharmaceutical
firms, $V_g\approx 0.019$ for manufacturing firms, and $V_g\approx 0.01$ for
products. After rescaling, the four PDFs can be fit by the same function. For
clarity, the pharmaceutical firms are offset by a factor of $10^2$,
manufacturing firms by a factor of $10^4$ and the pharmaceutical products by
a factor of $10^6$. Note that the data for pharmaceutical products extend
from $P_g(g)=1$ to $P_g(g)\approx 10^{-4}$ and the mismatch in the tail parts
is because $P_g(g)$ for large $g$ is mainly determined by the logarithmic
growth rates of units $\ln\eta$.}
\label{growth_dist}
\end{figure}

\begin{figure}[tbp]
\centering
\includegraphics[width=10cm,angle=-90]{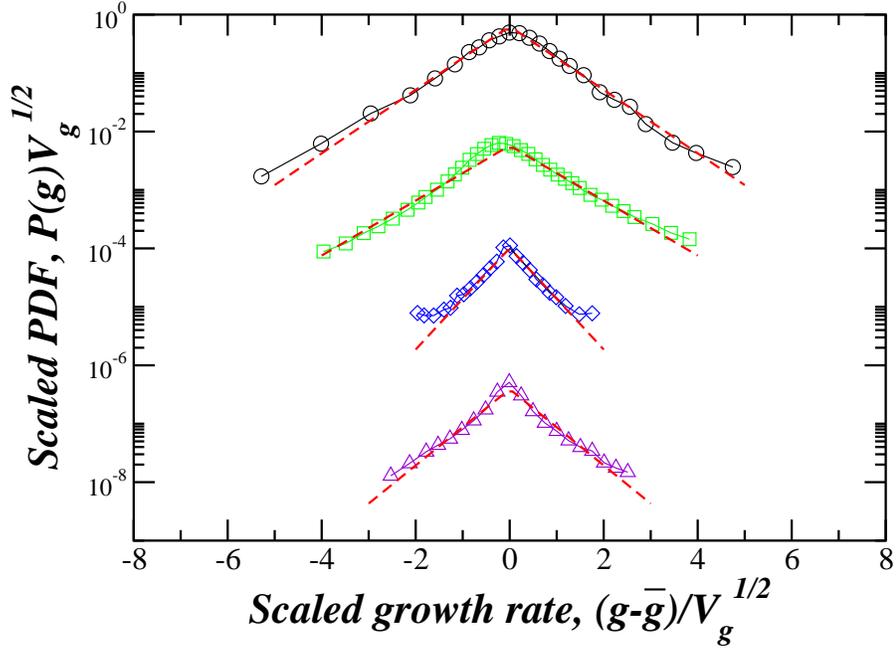}
\caption{Empirical tests of Eq.~(\protect\protect\ref{p_new_1}) for the {\it
central} part in the PDF $P(g)$ of growth rates rescaled by
$\sqrt{V_g}$. Shown are 4 symbols: country GDP ($\bigcirc$),
pharmaceutical firms ($\square$), manufacturing firms ($\Diamond$), and
pharmaceutical products ($\bigtriangleup$). The shape of central parts for
all four levels of aggregation can be well fit by a Laplace distribution
(dashed lines). Note that Laplace distribution can fit $P_g(g)$ only over a
restricted range, from $P_g(g) = 1$ to $P_g(g)\approx 10^{-1}$. }
\label{body}
\end{figure}

\begin{figure}[tbp]
\centering
\includegraphics[width=10cm,angle=-90]{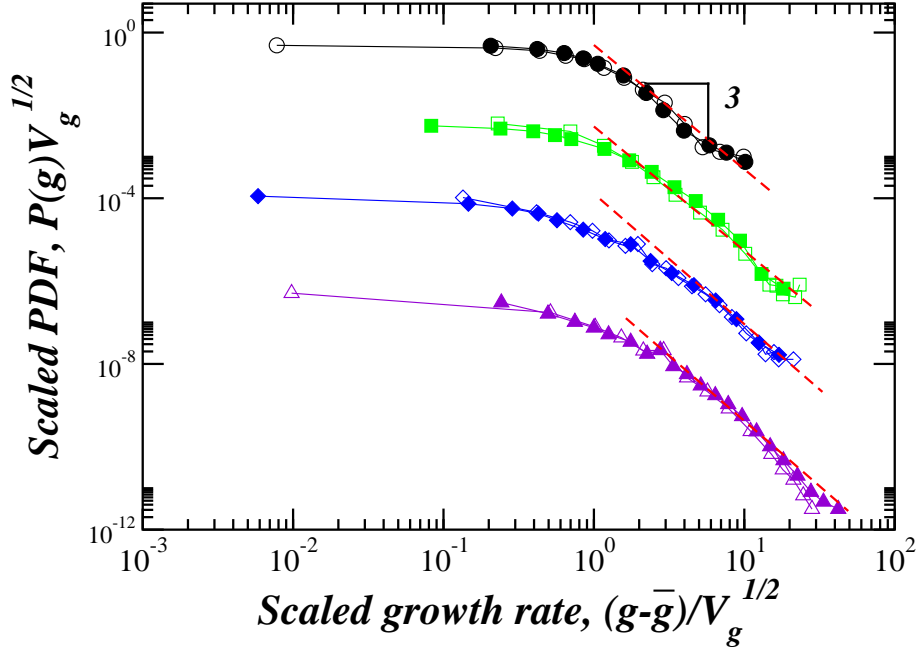}
\caption{Empirical tests of Eq.~(\protect\ref{p_new_1}) for the {\it tail}
parts of the PDF of growth rates rescaled by $\sqrt{V_g}$. The asymptotic
behavior of $g$ at any level of aggregation can be well approximated by power
laws with exponents $\protect\zeta\approx 3$ (dashed lines). The symbols are
as follows: Country GDP (left tail: $\bigcirc$, right tail: $\bullet$),
pharmaceutical firms (left tail: $\square$, right tail: $\blacksquare$),
manufacturing firms (left tail: $\Diamond$, right tail: $\blacklozenge$),
pharmaceutical products (left tail: $\bigtriangleup$, right tail:
$\blacktriangle$).} \label{tails}
\end{figure}

Fig.~\ref{growth_dist} shows that the growth distributions of countries,
firms, and products are well fitted by the distribution in
Eq.~(\protect\ref{p_new_1}) with different values of $V_g$.  Indeed, growth
distributions at any level of aggregation depict marked departures from a
Gaussian shape. Moreover, even if the $P_g(g)$ of GDP can be approximated by a
Laplace distribution, the $P_g(g)$ of firms and products are clearly more
leptokurtic than Laplace. Based on our model, the growth distribution is
Laplace in the body, with power-law tails. In fact, Fig.~\ref{body} show that
the central body part of the growth rate distributions at any level of
aggregation is well approximated by a double exponential
fit. Fig.~\ref{tails} reveals that the asymptotic behaviors of $g$ at any
level of aggregation can be well fitted by power-law with an exponent
$\zeta=3$.

Our analysis in Sec.~II predicts that the power law regime of $P_g(g)$ may
vary depending on the behavior of $P(K)$ for $K\to 1$, and the distribution
of the growth rates of units. In case of PHID, for which $P(1)\gg P(2) \gg
P(3) \ldots$ the growth rate distribution of firms must be almost the same as
the growth rate distribution of products, as we stated in Sec.~II. Hence the
power law wings of $P_g(g)$ for firms originate on the level of
products. Because PHID does not contain information on the subunits of products
we can not test our prediction directly, but we can hypothesize that the
distribution of the product subunits (number of customers or shipping ways)
is less dominated by small $K$, but has a sufficiently wide power law regime
due to the influx of new products. These rather plausible assumptions are
sufficient to explain the shape of the distribution of the product growth
rates, which is well described by Eq.~(\protect\ref{p_new_1}).

The PHID database allows us to test the empirical conditional distribution
$P_{g}(g|K)$ and the dependence of its variance $\sigma^2$ on $K$, where $K$
is the number of products. We find that $\sigma \sim K^{-0.28}$, which is
significantly smaller than $1/\sqrt{K}$ behavior. This result does not imply
correlations among product growth rates on the firm
level~\cite{DeFabritiis03}, but can be explained by the fact that for skewed
distributions of product sizes $P_{\xi}(\xi)$ characterized by large
$V_{\xi}$, the convergence of $P_g(g|K)$ to its Gaussian limit
Eq.~(\ref{P_g_gk}) is slow and the growth rates of the firms are determined
by the growth of the few large products. Using the empirical values for the
PHID $\mu_{\xi}=3.44$, $V_{\xi}=5.13$, $\mu_{\eta}=0.016$, $V_{\eta}=0.36$
and assuming lognormality of the distributions $P_{\xi}(\xi)$ and
$P_{\eta}(\eta)$ we find that the behavior of $\sigma$ can be well
approximated by a power law $\sigma \sim K^{-0.20}$ for $K<10^3$. For this
set of parameters, the convergence of $P_g(g|K)$ to a Gaussian distribution
takes place only for $K>10^5$. This result is consistent with the
observations of the power law relationship between firm size and growth rate
variance reported earlier~\cite{Stanley96,Amaral97,Sergey_II,Matia05}.
\section{Discussion}
Business firms grow by increasing their scale and scope. The scope of a firm
is given by the number of its products. The scale of a firm is given by the
size of its products. A firm like Microsoft gets few big products while Amazon
sells a huge variety of goods, each of small size in terms of sales. In this
article we argue that both mechanisms of growth are proportional. The number of
products a firm can successfully launch is proportional to the number of
products it has already commercialized. Once a product has been launched its
success depends on the number of customers who buy it and the price they are
willing to pay. To a large extent, if products are different enough, the
success of a product is independent from other products commercialized by the
same company. Hence, the sales of products can be modeled as independent
stochastic processes. Moreover, sometimes, new products are commercialized by
new companies. As a result, small companies with few products can experience
sudden jerks of growth due to the successful launch of a new product.

In this article, we find that the empirical distribution of firm growth rates
exhibits a central part which is distributed according to a Laplace
distribution and power-law wings $P_g(g)\thicksim g^{-\zeta }$ where
$\zeta=3$. If the distribution of number of units $K$ is dominated by single
unit classes, the tails of firm growth rate are primarily due to smaller
firms composed of one or few products. The Laplace center of the distribution
is shaped by big multiproduct firms. We find that the shape of the
distribution of firm growth is almost the same in presence of a small entry
rate and with zero entry. We also find that the model's predictions are
accurate also in the case of product growth rates, which implies that
products can be considered as composed by elementary sale units, which evolve
according to a random multiplicative process \cite{Steindl65}. Although there
are several plausible explanations for the Laplace body of the distribution
~\cite{Amaral97,Kotz01}, the power law decay of the tails has not previously
been observed. We introduce a simple and general model that accounts for both
the central part and the tails of the distribution. The shape of the business
growth rate distribution is due to the proportional growth of both the number
and the size of the constituent units in the class. This result holds in the
case of an {\em open} economy (with entry of new firms) as well as in the
case of a {\em closed} economy (with no entry of new firms).

\subsection*{Acknowledgment}
We thank S.~Havlin, J.~Nagler and F.~Wang for helpful discussions and
suggestions. We thank NSF and Merck Foundation (EPRIS Program) for financial
support.



\newpage

\appendix
\renewcommand{\theequation}{A\arabic{equation}}
\setcounter{equation}{0} 
{\noindent \textbf{\large Appendix A: The distribution of units in old and
new classes}}

Assume that at the beginning there are $N(0)$ classes with $n(0)$
units. Because at every time step one unit is added to the system and a new
class is added with probability $b$, at moment $t$ there are $n(t)= n(0) + t$
units and $N(t)=N(0)+bt$ classes, among which there are $bt$ new classes with
$n_{new}$ units and $N(0)$ old classes with $n_{old}$ units, such that
$n_{old} + n_{new} = n(0) + t$.

Because of the preferential attachment assumption, we have
\begin{eqnarray}
\frac{dn_{new}}{dt} & = & b + (1-b) \frac{n_{new}}{n(0) + t}, \\
\frac{dn_{old}}{dt} & = & (1-b) \frac{n_{old}}{n(0)+t}.
\end{eqnarray}
Solving the second differential equation and taking into account initial
condition $n_{old}(0)=n(0)$, we obtain
\begin{eqnarray}
n_{old}(t) & = & (n(0) + t)^{1-b} \,\, n(0)^b.
\end{eqnarray}
Analogously, the number of units at time $t$ in the classes existing at time
$t_0$ is
\begin{eqnarray*}
n_{e}(t_0,t) & = & (n(0) + t)^{1-b}(n(0) + t_0)^b,
\end{eqnarray*}
where the subscript `e' means ``existing''. The average number of units in
old classes is
\begin{eqnarray}
K(t) = \frac{ n_{old}(t) }{N(0)} = \frac{ (n(0) + t)^{1-b} }{N(0)}\,\, n(0)^b.
\label{mean_K_old}
\end{eqnarray}

It is known that~\cite{Cox} for $t\to\infty$ the preferential attachment
model converges to the exponential distribution:
\begin{equation}
P_{app}(K) \approx \exp(-K/K(t))/K(t).
\label{e_Cox}
\end{equation}
Thus, we obtain
\begin{eqnarray}
P_{old}(K) \approx \frac{N(0)}{(n(0) + t)^{1-b} n(0)^b}\,\,\,\exp\left( - \frac{K\,N(0)}{(n(0) +
t)^{1-b}n(0)^b}\right), \label{p_old_class}
\end{eqnarray}
and the part of $P(K)$ of old classes is
\begin{eqnarray}
\tilde{P}_{old}(K) \approx P_{old}(K) \frac{N(0)}{N(0)+t}.
\label{p_old_K_app}
\end{eqnarray}

The number of units in the classes that appear at $t_0$ is $b\,dt$ and the
number of these classes is $b\,dt$. Because the probability that a class
captures a new unit is proportional to the number of units it has already
gotten at time $t$, the number of units in the classes that appear at time
$t_0$ is
\begin{eqnarray*}
n_{new}(t_0,t) = n_{e}(t_0,t) \cdot \frac{bdt}{n(0) + t_0}.
\end{eqnarray*}

The average number of units in these classes is
$K(t_0,t)=n_{new}(t_0,t)/b\,dt=(n(0) + t)^{1-b}/(n(0)+t_0)^{1-b}$. Assuming
that the distribution of units in these classes is given by a continuous
approximation in Eq.~(\ref{e_Cox}):
\begin{eqnarray}
P_{new}(K) \approx \frac{1}{K(t_0,t)}\,\exp\left(-K/K(t_0,t) \right)
\end{eqnarray}
Thus, their contribution to the total distribution is
\begin{eqnarray*}
\frac{b\,dt_0}{N(0) + b\,t}\,\frac{1}{K(t_0,t)}\,\exp\left(-K/K(t_0,t) \right)
\end{eqnarray*}
The contribution of all new classes to the distribution $P(K)$ is
\begin{eqnarray}
\tilde{P}_{new}(K) \approx \frac{b}{N(0) + b\,t}\int_0^t
\frac{1}{K(t_0,t)}\,\exp\left(-K/K(t_0,t) \right)\, dt_0.
\end{eqnarray}
If we let $y = K/K(t_0,t)$, then
\begin{eqnarray}  \label{new}
\tilde{P}_{new}(K) & \approx & \frac{b}{1-b}\,\, K^{\left(-\frac{1}{1-b} -
1\right)}\, \frac{n(0)+t}{N(0)+bt}\,\, \int^K_{K\left(\frac{n(0)}{n(0)+t}
\right)^{1-b}} e^{-y} \,\,y^{\frac{1}{1-b}} \,\,dy.
\label{tilde_p_new_K}
\end{eqnarray}

Note that Eq.~(\ref{p_old_K_app}) and Eq.~(\ref{tilde_p_new_K}) are not exact
solutions but continuous approximations which assume $K$ is a real
number. Now we investigate the distribution in Eq.~(\protect\ref{new}).

1. At fixed $K$ when $t \rightarrow \infty$, the low limit of integration in
   Eq.~(\ref{tilde_p_new_K}) goes to zero and we have
\begin{eqnarray}
P_{new}(K) = \frac{K^{-1-\frac{1}{1-b}}}{1-b}\int_0^K
e^{-y}y^{\frac{1}{1-b}}dy.
\end{eqnarray}
As $K \rightarrow \infty$,
\begin{eqnarray}
P_{new}(K) & = & K^{-1-\frac{1}{b}} \left( \frac{1}{1-b}\right)
\,\,\Gamma\left(1+\frac{1}{1-b}\right).
\end{eqnarray}
As $K\rightarrow 0$,
\begin{eqnarray}
P_{new}(K) & = & \frac{1}{1-b}\,\, K^{\left(-\frac{1}{1-b} - 1\right)}\,\,
\frac{K^{\left(1+\frac{1}{1-b}\right)}}{1+\frac{1}{1-b}}  = \frac{1}{2-b}.
\end{eqnarray}
2. At fixed $t$ when $K \rightarrow \infty$, we use the partial integration to
   evaluate the incomplete $\Gamma$ function:
\begin{eqnarray*}
\,\,\,\,\,\, \int_x^{\infty}e^{-y}\,\,y^{\alpha}\,\,dy & = &
-e^{-y}\,\,y^\alpha |_x^\infty + \alpha\int_x^\infty
e^{-y}\,\,y^{\alpha-1}\,\,dy \approx e^{-x}\,\,x^{\alpha}.
\end{eqnarray*}
Therefore, from Eq.~(\ref{tilde_p_new_K}) we obtain
\begin{eqnarray}
P_{new}(K) &\approx & \frac{n(0)+t}{N(0)+bt}\,\,\frac{b}{1-b} \,\,K^{-\frac{1}{1-b}
  -1} \int^\infty_{K\left(\frac{n(0)}{n(0)+t} \right)^{1-b}} \,\,e^{-y}\,\,y^{
  \frac{1}{1-b}}\,\,dy, \notag \\ &=&
\frac{n(0)}{N(0)+bt}\,\,\frac{b}{1-b}\,\,\frac{1}{K}\,\,\exp\left(-K\left(
\frac{ n(0)}{n(0)+t}\right)^{1-b}\right),
\end{eqnarray}
which always decays faster than Eq.~(\ref{p_old_class}).


\medskip\medskip
{\noindent\textbf{\large Appendix B: Calculation of the growth distribution
of classes $P(g)$}}

\medskip\medskip
Let us assume both the size and growth of units ($\xi_i$ and $\eta_i$
respectively) are distributed lognormally
\begin{equation}
p(\xi_i)={\frac{1}{\sqrt{2\pi V_\xi}}}\,\,{\frac{1}{\xi_i}}
\,\,\exp\left(-(\ln\xi_i-m_\xi)^2/2V_\xi\right),
\end{equation}
\begin{equation}
p(\eta_i)={\frac{1}{\sqrt{2\pi V_\eta}}}\,\,{\frac{1}{\eta_i}}
\,\,\exp\left(-(\ln\eta_i-m_\eta)^2/2V_\eta\right).
\end{equation}
If units grow according to a multiplicative process, the size of units $
\xi_i^{\prime}=\xi_i\eta_i$ is distributed lognormally with $
V_{\xi^{\prime}}=V_\xi+V_\eta$ and $m_{\xi^{\prime}}=m_\xi + m_\eta$.

The $n^{\mbox{\scriptsize th}}$ moment of the variable $x$
distributed lognormally is given by
\begin{eqnarray}
\mu_x(n) &=& \int_0^\infty{\frac{1}{\sqrt{2\pi
V}}}\,{\frac{x^n}{x}}\,dx\,\exp\left(-(\ln x-m)^2/2V\right) \,\,=\,\,
\exp\left(nm_x+n^2V_x/2\right).
\label{mun}
\end{eqnarray}
Thus, its mean is $\mu_x \equiv \mu_x(n=1)= \exp(m_x+V_x/2)$ and its variance is
$\sigma_x^2\equiv \mu_{2}-\mu_{1}^2=\mu_{1}^2\,(\exp(V_x)-1)$.

Let us now find the distribution of $g$ growth rate of classes. It is defined
as
\begin{equation}
g\equiv \ln{\frac{S(t+1)}{S(t)}}=\ln\sum_{i=1}^K\xi_i^{\prime}-\ln\sum_{i=1}^K\xi_i.
\end{equation}
Here we neglect the influx of new units. According to the central limit
theorem, the sum of $K$ independent random variables with mean
$\mu_{\xi}\equiv \mu_{\xi}(1)$ and finite variance $\sigma_{\xi}^2$ is
\begin{equation}
\sum_{i=1}^K\xi_i=K\mu_\xi+\sqrt K\nu_K,
\end{equation}
where $\nu_K$ is the random variable with the distribution converging to
Gaussian
\begin{equation}
\lim_{K\to\infty}P(\nu_K)\to{\frac{1}{\sqrt{2\pi\sigma_\xi^2}}}\,\,
\exp\left(-\nu_K^2/2\sigma_\xi^2\right).
\end{equation}
Because $\ln \mu_{\eta}=m_\eta+V_\eta/2$ and $\ln
\mu_{\xi^{\prime}}=\ln \mu_{\xi} + \ln \mu_{\eta}$ we have
\begin{eqnarray}
g\equiv \ln S(t+1)-\ln S(t) &=& \ln(K\mu_{\xi^{\prime}})+{\frac{
\nu_K^{\prime}}{\sqrt K\mu_{\xi^{\prime}}}}- \ln(K\mu_\xi) - {\frac{\nu_K}{
\sqrt K\mu_{\xi}}}, \notag \\ &=&
m_\eta+{\frac{V_{\eta}}{2}}+{\frac{\nu_K^{\prime}\mu_\xi-\nu_K\mu_{\xi^{\prime}}}{\sqrt
K\mu_\xi\mu_{\xi^{\prime}}}}.
\label{g_m_V}
\end{eqnarray}
For large $K$ the last term in Eq.~(\ref{g_m_V}) is the difference of two
Gaussian variables and that is a Gaussian variable itself.
\begin{eqnarray}
\ln\mu_{\xi^{\prime}} = m_{\xi^{\prime}}+V_{\xi^{\prime}}/2 =
\ln\mu_\xi+\ln\mu_\eta,
\end{eqnarray}
where $\ln\mu_{\eta}=m_\eta+V_\eta/2$ is the average growth rate. To find the distribution
of $g$ we must find its mean and variance. In order to do this, we rewrite
\begin{equation*}
\frac{\nu_K^{\prime}}{\sqrt K\,\mu_{\xi^{\prime}}} = \frac{
\sum_{i=1}^K(\xi_i^{\prime}-\mu_{\xi^{\prime}})}{K\,\mu_{\xi^{\prime}}},
\end{equation*}
and
\begin{equation*}
\frac{\nu_K}{\sqrt K\,\mu_\xi}=\frac{\sum_{i=1}^K(\xi_i - \mu_\xi)}{
K\,\mu_\xi}.
\end{equation*}
Thus
\begin{eqnarray}
g &=&
m_\eta+{\frac{V_\eta}{2}}\,+\,{\frac{\sum_{i=1}^K\xi_i(\eta_i\mu_\xi-\mu_{\xi^{\prime}})}{K\mu_\xi\mu_{\xi^{\prime}}}},
\notag \\ &=& m_\eta+{\frac{V_\eta}{2}}+
{\frac{\sum_{i=1}^K\xi_i(\eta_i-\mu_\eta)}{K\mu_{\xi^{\prime}}}}.  
\label{g}
\end{eqnarray}

Because $\mu_{\xi^{\prime}} = \mu_{\xi} \mu_{\eta}$, the average of each term
in the sum is $\mu_{\xi^{\prime}}-\mu_\xi\,\mu_\eta=0$. The variance of each
term in the sum is $\langle(\xi_i\,\eta_i)^2\rangle-\langle
2\xi_i^2\,\eta_i\,\mu_\eta\rangle +\langle\xi_i^2\,\mu_{\eta}^2\rangle$ where
$\xi_i\eta_i$, $\xi_i^2\eta_i$ and $\xi_i^2$ are all lognormal independent
random variables. Particularly, $(\xi_i\eta_i)^2$ is lognormal with $V =
4V_\eta+4V_\xi$ and $m=2m_\eta+2m_\xi$; $\xi_i^2\eta_i$ is lognormal with $V
= 4V_\xi+V_\eta$ and $m = 2m_\xi+m_\eta$; $\xi_i^2$ is lognormal with $V =
4V_\xi$ and $m = 2m_\xi$. Using Eq.~(\protect\ref{mun}) and
Eq.~(\protect\ref{g})
\begin{subequations}
\begin{align}
\langle(\xi_i\eta_i)^2\rangle &=
\exp(2m_\eta+2m_\xi+2V_\eta+2V_\xi),\label{mean_all_a} \\
\langle\xi_i^2\eta_i\rangle &=
\exp(m_\eta+2m_\xi+2V_\xi+V_\eta/2),\label{mean_all_b} \\
\langle\xi_i^2\rangle &= \exp(2m_\xi+2V_\xi).
\label{mean_all_c}
\end{align}
\end{subequations}
Collecting all terms in Eqs.~(\ref{mean_all_a}-\ref{mean_all_c}) together and
using Eq.~(\ref{g}) we can find the variance of $g$:
\begin{eqnarray}
\sigma_g^2 &=&
\frac{K\,\exp(2m_\xi+2V_\xi+2m_\eta+V_\eta)(\exp(V_\eta)-1)}{K^2\exp(2m_\xi+V_\xi+2m_\eta+V_\eta)},
\notag \\ &=& \frac{1}{K}\exp(V_\xi)\,(\exp(V_\eta)-1).
\end{eqnarray}

Therefore, for large $K$, $g$ has a Gaussian distribution
\begin{equation}
P(g|K)={\frac{\sqrt K}{\sqrt{2\pi V}}}\,\exp\left(-\frac{(g-m)^2K}{2V}\right),
\label{P_g_large_K}
\end{equation}
where $m = m_\eta+ V_\eta/2$, $V= \exp(V_\xi)(\exp(V_\eta)-1)$
and $\mu_\eta = \exp(m_\eta+V_\eta/2)$.

The distribution of the growth rate of the old classes can be found by
Eq.~(\ref{P_g_g_sum}) in the text. In order to find a close form approximation,
we replace the summation in Eq.~(\ref{P_g_g_sum}) by integration and replace
the distributions $P(K)$ by Eq.~(\ref{p_old_class}) and $P(g)$ by the
Eq.~(\ref{P_g_large_K}) assuming $m=0$:
\begin{eqnarray}
P_{old}(g) & \approx & {\frac{1 }{\sqrt{2\pi V}}}\int_0^\infty{\frac{1}{K(t)}}\, \exp(
\frac{-K}{K(t)}) \exp(-\frac{g^2\,K}{2\,V})\sqrt{K}\,\,dK, \notag \\ &=&
\frac{\sqrt{K(t)}}{2\,\sqrt{2\,V}} \,\left(
1+\frac{K(t)}{2V}\,g^2\right)^{-\frac{ 3}{2}}, \label{e.*}
\end{eqnarray}
where $K(t)$ is the average number of units in the old classes (see
Eq.~(\ref{mean_K_old})). This distribution decays as $1/g^3$ and thus does
not have finite variance.  In fact, we approximate the distribution of number
of units in the old classes by a continuous function $\exp(-K/K(t))/K(t)$,
while in reality it is a discrete distribution
\begin{equation}
P_{\mbox{\scriptsize old}}(K)=\lambda^K\left({\frac{1}{\lambda}}-1\right),
\end{equation}
where $\lambda=\exp(-1/K(t))$. The corrected distribution of growth rates is
then given by the sum
\begin{equation}
P_{\mbox{\scriptsize old}}(g) \approx {\frac{1}{\sqrt{2\pi V}}}{\frac{1-\lambda}{
\lambda}}\sum_{K=1}^\infty\lambda^K\,\sqrt{K}\,\exp(-g^2 K/2V).
\end{equation}
The slowest decaying term is
\begin{equation}
{\frac{1}{\sqrt{2\pi V}}}\,\,(1-\lambda)\,\,\exp(-g^2/2V),  \label{e.**}
\end{equation}
which describes the behavior of the distribution when $g\to\infty$.
Thus there is a crossover from Eq.~(\ref{e.*}) to Eq.~(\ref{e.**}) when
$g\approx\sqrt{2V}$.

For the new classes, when $t\to\infty$ the distribution of number of units
is approximated by
\begin{equation}
P_{\mbox{\scriptsize new}}(K) \approx {\frac{1}{1-b}}K^{-1-\frac{1}{1-b}
}\,\int_0^K\,y^{\frac{1}{1-b}}\,e^{-y}\,\,dy.  \label{e.***}
\end{equation}

Again replacing summation in Eq.~(\ref{P_g_g_sum}) in the text by integration
and $P(g|K)$ by Eq.~(\ref{P_g_large_K}) and after the switching the order
of integration we have:
\begin{equation}
P_{new}(g) \approx {\frac{1}{1-b}}\,{\frac{1}{\sqrt{2\pi
V}}}\int_0^\infty\,\exp(-y) \,y^{\frac{1}{1-b}}\,dy\,\int_y^\infty\,
\exp(-g^2\,K/2V)\,K^{(-\frac{1}{2}-\frac{1}{1-b}) }\,dK.  \label{e.****}
\end{equation}

As $g\to\infty$, we can evaluate the second integral in Eq.~(\ref{e.****}) by
partial integration:
\begin{eqnarray}
P_{new}(g) & \approx & \frac{1}{1-b}\int_0^\infty \frac{1}{\sqrt{2\pi V}} \,\, \frac{
2V}{g^2} \,\, y^{-\frac{1}{1-b} - \frac{1}{2}} \,\, y^{\frac{1}{1-b}} \,\,
\exp(-y)\,\,\exp(-y\,g^2/2V)\,\,dy, \notag \\ &=& \frac{1}{1-b}\,
\frac{1}{\sqrt{2\pi V}}\,\frac{2V}{g^2} \,\,\frac{1}{
\sqrt{g^2/2V+1}}\,\, \sqrt{\pi} \sim \frac{1}{g^3}.
\end{eqnarray}

We compute the first derivative of the distribution (\ref{e.****}) by
differentiating the integrand in the second integral with respect to $g$.
The second integral converges as $y\to 0$, and we find the behavior of the
derivative for $g\to 0$ by the substitution $K^{\prime}=K g^2/(2V)$. As $g\to
0$, the derivative behaves as $g\cdot g^{2[-(3/2)+1/(1-b)]}\sim
g^{1/(1-b)-2}$, which means that the function itself behaves as $%
C_2-C_1|g|^{2b/(1-b)+1}$, where $C_2$ and $C_1$ are positive constants. For
small $b$ this behavior is similar to the behavior of a Laplace distribution
with variance $V$:
$\exp(-\sqrt{2}|g|/\sqrt{V})/\sqrt{2V}=1/\sqrt{2V}-|g|/V$.

When $b\to 0$, Eq.~(\ref{e.****}) can be simplified:
\begin{eqnarray}
P_{new}(g)|_{b\to 0} & \approx & \frac{1}{\sqrt{2\pi V}} \, \int_0^\infty
K^{-3/2}\,\exp(-K\,g^2/2\,V)\,dK\,\int_0^K\,\exp(-y)\,y\,\,dy,  \notag \\
& \approx & \frac{1}{\sqrt{2\,V}}\,\,\left(-\frac{1}{\sqrt{1+ g^2/2\,V}} +
\frac{2}{|g|/\sqrt{2\,V} + \sqrt{g^2/2\,V+1 }} \right).
\notag
\end{eqnarray}
Finally we find
\begin{equation}  \label{p_new}
P_{\mbox{\scriptsize new}}(g)|_{b\to 0} \approx \frac{2V}{\sqrt{g^2+2V}\,(|g|+\sqrt{g^2+2V})^2}.
\end{equation}
which behaves for $g\to 0$ as $1/\sqrt{2V}-|g|/V$ and for $g\to\infty$ as
$V/(2g^3)$. Thus the distribution is well approximated by a Laplace
distribution in the body with power-law tails.

Because of the discrete nature of the distribution of the number of units,
when $g\gg\sqrt{2V}$ the behavior for $g\to\infty$ is dominated by
$const\cdot\, \exp(-g^2/2V)$.

\newpage


\eject

\eject

\eject

\end{document}